\documentclass[9pt,twocolumn,twoside]{opticajnl}
\journal{opticajournal} 

\setboolean{shortarticle}{false}
 
\usepackage{natbib}
\usepackage{bm}
\usepackage{url}
\usepackage{pifont}
\usepackage{array}
\usepackage{tensor}
\usepackage{amsthm}
\usepackage{amsmath}
\usepackage[makeroom]{cancel}
\usepackage{amssymb}
\usepackage{amsfonts}
\usepackage{mathrsfs}
\usepackage{mathtools}
\usepackage{nicefrac}
\usepackage{graphicx}
\usepackage{physics}
\usepackage{color,soul}
\usepackage[normalem]{ulem}
\usepackage{color}
\usepackage{placeins}
\usepackage{float}
\usepackage{hyperref}
\usepackage{tabularx,colortbl}
\usepackage{amsmath}
\usepackage{psfrag}
\usepackage{pstool}
\usepackage{appendix}
\usepackage{lipsum}

\newcommand{\tx}{\text}

\newcommand{\nb}{\noindent$\bullet$~}

\usepackage{lineno}

\title{Orthogonal Analysis of Space-Time Crystals}
\author[1,*]{Zoé-Lise Deck-Léger}
\author[2]{Amir Bahrami}
\author[3,2]{Zhiyu Li}
\author[2]{Christophe Caloz}

\affil[1]{Polytechnique Montréal, 2500, ch. de Polytechnique,
Montreal, H3T 1J4, Quebec, Canada}
 \affil[2]{KU Leuven, Kasteelpark Arenberg 10 Box 2440
3001 Leuven,
Belgium}
\affil[3]{Xi'an Jiaotong University, West Xianning Road 28, Box 710049, Xi'an, Shaanxi, China}

\affil[*]{zoe-lise.deck-leger@polymtl.ca}

\begin{abstract}
This paper presents a space-time-wise orthogonal analysis of space-time crystals. This analysis provides a solution consisting of a pair of explicit parametric equations that result from a separate application of the Bloch-Floquet theorem in the (orthogonal) directions of space and time. Compared to previous approaches, this solution offers the benefits of greater simplicity, clearer emphasis on space-time duality and deeper physical insight.
\end{abstract}

\setboolean{displaycopyright}{false} 

\begin{document}

\maketitle

The emergence of photonic crystals~\cite{yarivbook1984,skorobogatiy2009fundamentals,joannopoulos2011photonic} in the late 1980s as the optical counterparts of solid-state crystals~\cite{ashcroft}  represented a major development in the field of electromagnetics. Recently, these crystals have gained even greater interest with the \emph{addition of temporal modulation} to their spatial modulation within the framework of research on space-time metamaterials~\cite{caloz2019spacetime1,caloz2019spacetime2,caloz2022GSTEM}. The resulting \emph{space-time crystals} have already been demonstrated to support a diversity of novel effects, such as distributed parametric amplification~\cite{cullen1958travelling,lurie2006wave,pendry_gain_2021}, magnetless isolation~\cite{yu2009isolation,chamanara2017isolation}, space-time weighted averaging ``dragging''~\cite{deck2019uniform,caloz2022GSTEM}, beam deflection~\cite{huidobro2019fresnel} and beam curving~\cite{bahrami2022electrodynamics}.

The addition of the time dimension dramatically complicates the computation of the dispersion diagram (or photonic bandgap structure) of photonic crystals. An early attempt to address the related challenge has been the plane-wave expansion method reported in~\cite{cassedy1963p1,chutamir1972}. However, this approach results in transcendental equations that are intractable and offer no insight into the physics of the crystals. An a priori promising approach to resolve these issues might a priori seem to be the (spectral) frame-hopping technique~\cite{bladel1984}, using respectively stationary and instantaneous moving frames for the space-like and time-like regimes~\cite{deck2017superluminal,caloz2019spacetime2}, but this technique would in fact involve cumbersome bianisotropic parameters~\cite{Rontgen_1888,kong1968,caloz2022GSTEM} and complex space-time frequency coupling~\cite{caloz2019spacetime1,caloz2019spacetime2}. A \emph{partial} frame-hopping approach, applying periodicity in the direction of the moving frame but keeping quantities in the laboratory frame, was used in~\cite{biancalana2007dynamics,deck2019uniform} to remove the bianisotropic complication; however, this approach could not resolve the issue of coupling complexity.

We present here a \emph{space-time-wise orthogonal approach} of space-time crystals, completely conducted in the laboratory frame, that avoids both bianisotropy and space-time frequency coupling. This approach offers thus maximum simplicity. Furthermore, it unveils fundamental symmetries between the spatial and temporal frequencies, hence highlighting strong duality between the two, and provides deeper insight into the physics of the crystal.

\begin{figure}[h!]
\centering
\includegraphics[width=\columnwidth]{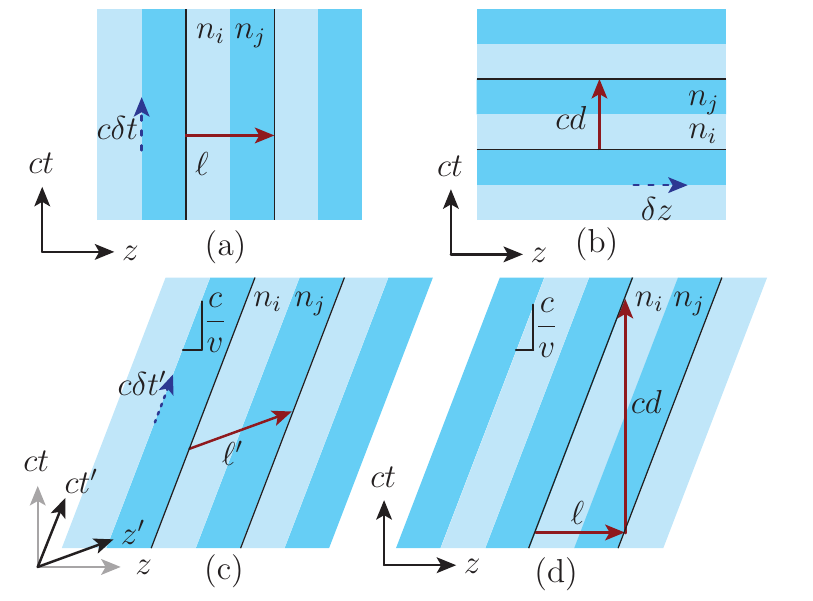}
\caption{Basic crystals, composed of layers of media with alternating refractive indices ($n_i$ and $n_j$), featuring discrete and continuous translation symmetries. (a)~Spatial crystal. (b)~Temporal crystal. (c)~Space-time crystal with discrete spatial translation in the moving frame. (d)~Space-time crystal with discrete spatial and temporal translations in the direct frame.}
\label{fig:structures}
\end{figure}

The spirit of the proposed solution may be best understood with the help of the basic crystals and related symmetries depicted in Fig.~\ref{fig:structures}. Figure~\ref{fig:structures}(a) shows a purely spatial crystal, with discrete translation symmetry, or periodicity (length $\ell$), in space, leading to the production of wavevector harmonics, and with continuous translation symmetry ($\delta t$) in time, corresponding to frequency conservation~\cite{skorobogatiy2009fundamentals}. Figure~\ref{fig:structures}(b) shows a purely temporal crystal (e.g.~\cite{galiffi_time_2022}), with discrete translation symmetry, or periodicity (duration $d$), in time, leading to the production of frequency harmonics, and with continuous translation symmetry ($\delta z$) in space, corresponding to momentum conservation~\cite{caloz2019spacetime2}. Figures~\ref{fig:structures}(c) and (d) show a space-time crystal~\cite{biancalana2007dynamics,deck2019uniform}, with modulation velocity $v$, under different symmetry perspectives: Figure~\ref{fig:structures}(c) highlights the moving-frame perspective~\cite{biancalana2007dynamics,deck2019uniform}, with continuous temporal and discrete spatial symmetries, which could be considered as the logical extension of the spatial crystal in a frame-hoping logic, whereas Fig.~\ref{fig:structures}(d) highlights the space-time-wise orthogonal perspective that will be taken in this paper, with discrete translation symmetries, or periodicities, in both space and time. Note that Figs.~\ref{fig:structures}(c) and (d) pertain to \emph{subluminal} ($v<c/n_{i,j}$, where $c/n_{i,j}$ is the speed of light in a medium with refractive index $n_{i,j}$) space-time crystals; the corresponding figures for \emph{superluminal} crystals ($v>c/n_{i,j}$) straightforwardly follow from subluminal-superluminal duality~\cite{deck2017superluminal}.

According to the elected strategy of orthogonal treatment of space and time, we shall proceed as follows in our analysis of the crystal: 1)~establish the field waveforms with their motion-shifted frequencies and motion-deflected propagation angles in each layer of the crystal; 2)~separately compute the corresponding spatial and temporal unit-cell [Fig.~\ref{fig:structures}(d)] transmission matrices; 3)~apply the Bloch-Floquet theorem to each of these matrices to obtain parametric relations for the Bloch-Floquet wavevector and frequency; 4)~combine these relations to plot the dispersion diagram and isofrequency curves of the crystal.

\begin{figure}[ht]
\centering
\includegraphics[width=1\columnwidth]{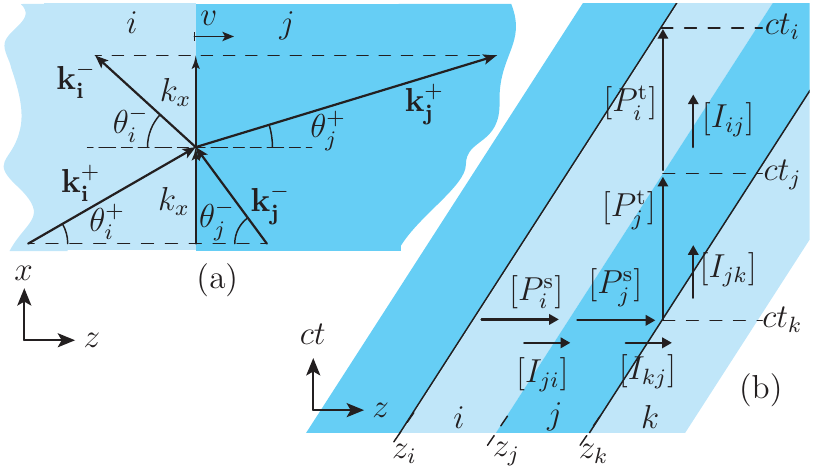}
\caption{Crystal unit-cell features. (a)~Frequency transitions and deflection angles at the interface between layers $i$ and $j$. (b)~Spatial and temporal unit-cell transfer matrices.}
\label{fig:plane_waves}
\end{figure}

Assuming propagation and scattering in the $x-z$ plane and s-polarization (the p-polarization case can be treated analogously), as illustrated in Fig.~\ref{fig:plane_waves}, the electric field waveform in the crystal may be written as the succession of individual layer waveforms, which read
\begin{subequations}\label{eq:efield_gen}
\begin{equation}\label{eq:efield}
E_{n}^\pm=A_{n}^\pm\tx{e}^{i(k_{x,n}^\pm x\pm k_{z,n}^\pm z-\omega_{n}^\pm t)},
\end{equation}
with
\begin{equation}\label{eq:kz_polar}
    k_{z,n}^\pm=\frac{n_n}{c}\omega_n^\pm \cos\theta_n^\pm, \quad k_{x,n}^\pm=\frac{n_n}{c}\omega_n^\pm \sin\theta_n^\pm,
\end{equation}
\end{subequations}
where the subscript $n$ refers to an arbitrary crystal layer and where the superscript signs correspond to propagation in the $(+z,+x)$ -- or forward -- and $(-z,+x)$ -- or backward -- directions, respectively.   

The motion-shifted frequencies $\omega_n^\pm$ and motion-deflected propagation angles $\theta_n^\pm$ in Eqs.~(\ref{eq:kz_polar}) may be found with the help of Fig.~\ref{fig:plane_waves}(a), which represents the corresponding wavevectors across two layers, labeled $i$ and $j$. The phases $\phi_n^\pm=k_{x,n}^\pm x\pm k_{z,n}^\pm z-\omega_{n}^\pm t$ ($n=i,j$) in~\eqref{eq:efield} must be conserved at the interface. Due to the absence of motion in the $x$ direction, the component $k_x$ of the wavevector is conserved at the crystal's interfaces and is hence unique across the crystal. In contrast, the frequencies, $\omega_n^\pm$, are not conserved, due to motion in the $z$ direction. Enforcing the continuity of the phases at $z=z_n+vt$, with $k_{z,n}^\pm$ given in~\eqref{eq:kz_polar} and $k_{x,n}^\pm=k_x$ yields~\cite{kunz1980plane}
\begin{equation}\label{eq:freq_relations}
\omega^\pm_n(\omega_i^+)=\frac{1-\beta n_i\cos\theta_i^+}{1\mp\beta n_n\cos\theta_n^\pm}\omega^+_i,
\end{equation}
where $\beta=v/c$ is the normalized modulation velocity. The corresponding deflection angles are then found by substituting $\omega_{n,i}^\pm$ from the second expression in~\eqref{eq:kz_polar} into~\eqref{eq:freq_relations}, and solving for $\cos\theta_n^\pm$, which yields~\cite{kunz1980plane} 
\begin{subequations}\label{eq:angles}
\begin{equation}
\cos\theta_n^\pm(\theta_i^+)=\pm\frac{ C^2\beta\pm D_n }{(1+C^2\beta^2)n_n},
  \end{equation}
with
\begin{equation}
    C=\frac{n_i\sin \theta_i^+}{1-\beta n_i\cos\theta_i^+}
\quad\text{and\quad}D_n=\sqrt{(1+C^2\beta^2)n_n^2-C^2}.
\end{equation}
\end{subequations}
Note that, at this point, all the quantities in Eqs.~(\ref{eq:efield_gen}) are expressed in terms of the parameters $\omega_i^+$ and $\theta_i^+$ via Eqs.~(\ref{eq:freq_relations}) and~(\ref{eq:angles}).

We can now compute the spatial and temporal transfer matrices of the unit cell. The spatial matrix, corresponding to the horizontal arrows in Fig.~\ref{fig:plane_waves}(b), $[M_{ki}^\tx{s}]$, relates the fields at the output of the unit cell, $E_k(z_k)$, to those at the input, $E_i(z_i)$, as
\begin{equation}\label{eq:spatial_uc}
\begin{bmatrix}
E_{k}^+\\E_{k}^-\end{bmatrix}_{z=z_k}=[M_{ki}^\tx{s}]\begin{bmatrix}E_{i}^+\\E_{i}^-\end{bmatrix}_{z=z_i}
=[I_{kj}][P_j^\tx{s}][I_{ji}][P_i^\tx{s}]\begin{bmatrix}E_{i}^+\\E_{i}^-\end{bmatrix}_{z=z_i},
\end{equation}
where $[M_{ki}^\tx{s}]$ is composed of interface matrices  $[I_{kj}]$ and $[I_{ji}]$ and propagation matrices $[P_j^\tx{s}]$ and $[P_i^\tx{s}]$, which are respectively defined by 
\begin{equation}\label{eq:spatial_propag}
\begin{bmatrix}E_{n+1}^+\\E_{n+1}^-\end{bmatrix}_{z_n}=[I_{n+1,n}]\begin{bmatrix}E_{n}^+\\E_{n}^-\end{bmatrix}_{z_{n}},\, 
\begin{bmatrix}E_{n}^+\\E_{n}^-\end{bmatrix}_{z_{n+1}}=[P_{n}^\tx{s}]\begin{bmatrix}E_{n}^+\\E_{n}^-\end{bmatrix}_{z_{n}}.
\end{equation}
The interface matrices are found by applying the boundary conditions in terms of the electric fields at the moving interface and read (see App.~\ref{app:Interface_Matrix_Derivation})
\begin{subequations}\label{eq:Iji}  
\begin{equation}
 [I_{mn}]
  =\frac{1}{2f_m^+f_m^-}\begin{bmatrix}
   f_n^+g_m^-+f_m^-g_n^+& -f_n^-g_m^-+f_m^-g_n^-\\
  -f_n^+g_m^++f_m^+g_n^+ & f_n^-g_m^++f_m^+g_n^-
 \end{bmatrix},
\end{equation}
with
\begin{equation}\label{eq:fgpm}
f_n^\pm=1\mp\beta n_n\cos\theta_n^\pm , \quad \text{and}\quad g_n^\pm=\frac{1}{\eta_n}\left(\cos\theta_n^\pm\mp\beta n_n\right),
\end{equation}
\end{subequations}
while the propagation matrices are obtained by inserting~\eqref{eq:efield} into the second equation of \eqref{eq:spatial_propag} with $z_{n+1}=z_n+\ell_n$, and read
\begin{equation}\label{eq:Ps}
  [P_n^\tx{s}]=\begin{bmatrix}
   e^{ik_{z,n}^+\ell_n}& 0\\
  0 & e^{-ik_{z,n}^-\ell_n}
 \end{bmatrix}.
\end{equation}

Similarly, the temporal matrix, following the vertical arrows in Fig.~\ref{fig:plane_waves}(b), $[M_{ik}^\tx{t}]$, relates the fields at the output of the unit cell, $E_i(t_i)$, to those at the input, $E_k(t_k)$, as
\begin{equation}\label{eq:temporal_uc}
\begin{bmatrix}E_{i}^+\\E_{i}^-\end{bmatrix}_{t=t_i}=[M_{ik}^\tx{t}]\begin{bmatrix}E_{k}^+\\E_{k}^-\end{bmatrix}_{t=t_k}=[P_i^\tx{t}][I_{ij}][P_j^\tx{t}][I_{jk}]\begin{bmatrix}E_{k}^+\\E_{k}^-\end{bmatrix}_{t=t_k},
\end{equation}
where the interface matrices correspond to those for the spatial unit-cell [Eqs.~\ref{eq:Iji}] with inverted indices, while the temporal propagation matrices reads
\begin{equation}\label{eq:Pt}
    [P_n^\tx{t}]=
   \begin{bmatrix}
       e^{i \omega^+_nd_n} & 0 \\
        0 & e^{i \omega_n^-d_n}
      \end{bmatrix}.
\end{equation} 

We can now proceed to the application of the Bloch-Floquet theorem in the spatial and temporal directions of the crystal to obtain the corresponding wavevectors and frequencies. In the spatial direction, the theorem states~\cite{bloch1929quantenmechanik,floquet1883equations}
\begin{equation}\label{eq:efield_bf}
E^\pm(z+\ell)=e^{\pm i k_{z}^\pm \ell}E^\pm(z),
\end{equation}
where $k_{z}^\pm$ are the sought-after Bloch-Floquet wavevectors. Successively substituting $E^\pm(z)=E^\pm_i(z_i)$ and $E^\pm(z+\ell)=E_k^\pm(z_k)$ from Fig.~\ref{fig:plane_waves}(b) and the first relation of \eqref{eq:spatial_uc} into this equation yields
\begin{equation}\label{eq:kz_matrix}
\begin{bmatrix}
   E_k^+\\
   E_k^-
\end{bmatrix}_{z_k}=\tx{e}^{\pm i k_z^\pm \ell}\begin{bmatrix}
   E_i^+\\
   E_i^-
\end{bmatrix}_{z_i}
=[M_{ki}^\tx{s}]\begin{bmatrix}
   E_i^+\\
   E_i^-
\end{bmatrix}_{z_k}.
\end{equation}
The second equality in \eqref{eq:kz_matrix} is an eigenvalue matrix equation, whose characteristic equation reads
\begin{subequations}
\begin{equation}\label{eq:exp_spat}
\tx{e}^{\pm 2ik_z^\pm \ell}-\tx{tr}[M_{ki}^\tx{s}]\tx{e}^{\pm ik_z^\pm \ell}+\tx{det}[M_{ki}^\tx{s}]=0,
\end{equation}
with, using~\eqref{eq:spatial_uc} with Eqs.~(\ref{eq:Iji}) and~(\ref{eq:Ps}), the explicit terms
\begin{equation}\label{eq:spatial_det}
   \tx{det}[M_{ki}^\tx{s}]=e^{i\Delta\phi^\tx{s}},
\end{equation}
and
\begin{equation}\label{eq:spatial_tr}
\tx{tr}[M_{ki}^\tx{s}]=2e^{i\Delta\phi^\tx{s}/2}\left(C_i^\tx{s}C_j^\tx{s}-AS_i^\tx{s}S_j^\tx{s}\right)=2e^{i\Delta\phi^\tx{s}/2}\tx{tr}[M_{ki,0}^\tx{s}],
\end{equation}
where 
$\Delta\phi^\tx{s}=\Delta\phi_i^\tx{s}+\Delta\phi_j^\tx{s}$, $C_n^\tx{s}=\cos(\Sigma\phi^\tx{s}_n/2)$, $S_n^\tx{s}=\sin(\Sigma\phi^\tx{s}_n/2)$,
\begin{equation}\label{eq:def_barphis}
\Delta \phi_n^\tx{s}=(k_{z,n}^+-k_{z,n}^-)\ell_n, \quad\Sigma{\phi}^\tx{s}_{n}=(k_{z,n}^++k_{z,n}^-)\ell_{n},
\end{equation}
and
\begin{equation}\label{eq:A}
A(\theta_i^+)=\frac{{f_j^+}^2{g_i^+}^2+{f_i^+}^2{g_j^+}^2}{2f_i^+f_j^+g_i^+g_j^+}.
\end{equation}
\end{subequations}
Multiplying now \eqref{eq:exp_spat} with \eqref{eq:spatial_det} by $\tx{e}^{-i\Delta\phi^\tx{s}}$ leads to
\begin{equation}\label{eq:exp_spat_2}
\tx{e}^{2i(\pm k_z^\pm \ell-\Delta \phi^\tx{s}/2)}-\tx{tr}[M_{ki,0}^\tx{s}]\tx{e}^{i(\pm k_z^\pm \ell-\Delta \phi^\tx{s}/2)}+1=0,
\end{equation}
which is a quadratic equation with solutions
\begin{equation}\label{eq:exp_spat_3}
\tx{e}^{\pm i( k_z^\pm \ell\mp\Delta\phi^\tx{s}/2)}=\frac{\tx{tr}[M_{ki,0}^\tx{s}]}{2}\pm i \sqrt{1-\left(\frac{\tx{tr}[M_{ki,0}^\tx{s}]}{2}\right)^2}.
\end{equation}
According to Euler's formula, $e^{\pm i\theta}=\cos\theta\pm i\sin\theta$, this relation is equivalent to $\cos( k_z^\pm\ell\mp\Delta\phi^{\tx{s}}/2)=\tx{tr}[M_{ki,0}^\tx{s}]/2$, whose inversion yields the explicit formula
\begin{equation}\label{eq:disp_kz}
k_z^\pm(\omega_i^+,\theta_i^+)=\frac{1}{\ell}\left[\acos\left(\frac{\tx{tr}[M_{ki,0}^\tx{s}]}{2}\right)\pm\Delta\phi^\tx{s}/2\right].
\end{equation}

Similarly, in the temporal direction, the Bloch-Floquet theorem states
\begin{equation}\label{eq:efield_bf_time}
E^\pm(t+d)=e^{i \omega^\pm d}E^\pm(t),
\end{equation}
where $\omega^\pm$ are the sought-after Bloch-Floquet frequencies. Successively substituting $E^\pm(t)=E^\pm_k(t_k)$ and $E^\pm(t+d)=E_i^\pm(t_i)$ from Fig.~\ref{fig:plane_waves}(b) and the first relation of \eqref{eq:temporal_uc} into this equation yields
\begin{equation}\label{eq:bf_time_matrix}
\begin{bmatrix}
   E_i^+\\
   E_i^-
\end{bmatrix}_{t_i}
=\tx{e}^{-i\omega^\pm d}
\begin{bmatrix}
   E_k^+\\
   E_k^-
\end{bmatrix}_{t_k}
=[M_{ij}^\tx{t}]\begin{bmatrix}
   E_k^+\\
   E_k^-
\end{bmatrix}_{t_k}.
\end{equation}
A treatment of this (second equality) eigenvalue matrix equation analogous to that applied to the spatial case straightforwardly leads to the explicit formula
\begin{equation}\label{eq:disp_w}
\pm\omega^\pm(\omega_i^+,\theta_i^+) =\frac{1}{d}\left( \acos(\frac{\tx{tr}[M_{ik,0}^\tx{t}]}{2})\pm\Sigma\phi^\tx{t}/2\right),
\end{equation}
where, from~\eqref{eq:temporal_uc} with Eqs.~(\ref{eq:Iji}) and~(\ref{eq:Pt}), 
\begin{equation}\label{eq:temp_det}
   \tx{det}[M_{ik}^\tx{t}]=e^{i\Sigma\phi^\tx{t}}
\end{equation}
and
\begin{equation}\label{eq:temp_tr}
   \tx{tr}[M_{ki}^\tx{t}]=e^{i\Sigma\phi^\tx{t}/2}\left(C_i^\tx{t}C_j^\tx{t}-AS_i^\tx{t}S_j^\tx{t}\right)=e^{i\Sigma\phi^\tx{t}/2}\tx{tr}[M_{ik,0}^\tx{t}],
\end{equation}
with $\Sigma\phi^\tx{t}=\Sigma\phi_i^\tx{t}+\Sigma\phi_j^\tx{t}$, $C_n=\cos(\Delta\phi^\tx{t}_n/2)$, $S_n=\sin(\Delta\phi^\tx{t}_n/2)$,
\begin{equation}\label{eq:def_barphit}
\Sigma{\phi}_n^\tx{t}=(\omega_n^++\omega_n^-) d_i, \quad\text{and}\quad \Delta \phi^\tx{t}_{n}=(\omega_n^+-\omega_n^-) d_{n},
\end{equation}
and with $A$ given in~\eqref{eq:A}.

Our orthogonal spatio-temporal treatment of the problem, illustrated in Fig.~\ref{fig:structures}(d), has thus led to a pair of \emph{separate} and~\emph{explicit} parametric (parameters $\omega_i^+$ and $\theta_i^+$) formulas for the Bloch-Floquet wavevectors and frequencies of the crystal, namely formulas~(\ref{eq:disp_kz}) and (\ref{eq:disp_w}), which represents the central result of the paper. These formulas offer the following benefits: \\
\nb Being separate and explicit, they offer maximal simplicity.
\nb Their duality, in terms of the substitutions $k_z^\pm\leftrightarrow\pm\omega^\pm$, $\ell\leftrightarrow d$, $\Delta\phi^\tx{s}\leftrightarrow\Sigma\phi^\tx{t}$ and $\Sigma\phi^\tx{s}\leftrightarrow\Delta\phi^\tx{t}$, reveals perfect spatio-temporal symmetry. \\
\nb Their $\acos$ function term, being multi- and complex-valued, provides physical insight into the spatio-temporal periodicity and band structure of 
the crystal. \\
\nb The non-zero value of their parameters $\Delta \phi^\tx{s}$ and $\Sigma\phi^\tx{t}$, due to different forward and backward features, is a direct manifestation and measure of nonreciprocity~\cite{rayleigh1885,Caloz_PRAp_10_2018}.

\begin{figure}[h!]
\centering
\includegraphics[width=0.7\columnwidth]{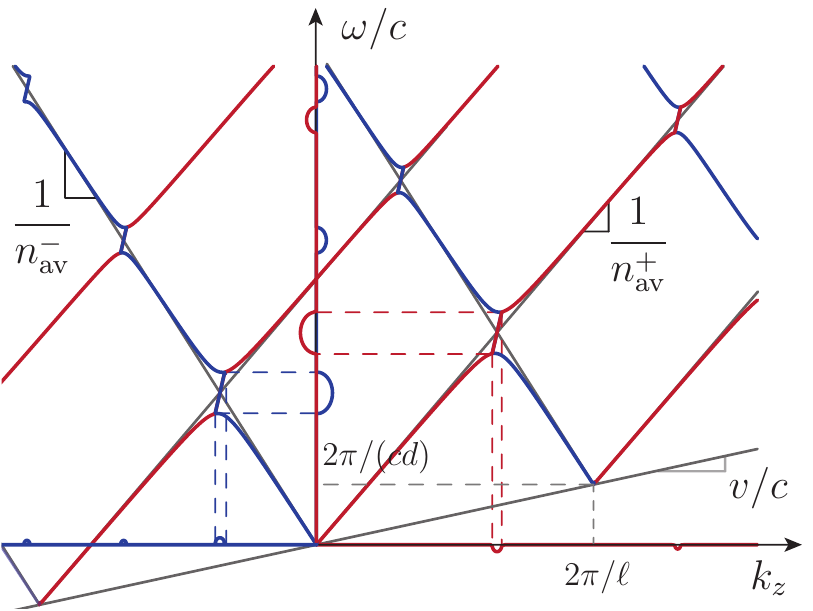}
\caption{Dispersion diagram for $\epsilon_1=1$, $\epsilon_2=1.5$ $\mu_1=\mu_2=1$, $\ell_1=\ell_2=\ell/2$, $v=0.2c$ and $\theta_i^+=40^\circ$, normalized to $\ell$ ($d=\ell /v$), with gray curves corresponding to the empty-lattice approximation found in App.~\ref{app:empty-lattice}}
\label{fig:Dispersion}
\end{figure}

Figure~\ref{fig:Dispersion} shows a typical dispersion diagram, corresponding the dispersion relation $\omega^\pm(k_z^\pm)$, which has been plotted by varying the parameter $\omega_i^+$ with parameter $\theta_i^+$ fixed in Eqs.~(\ref{eq:disp_kz}) and~(\ref{eq:disp_w}). Note the double average slope and gap position asymmetry, or nonreciprocity, in that diagram. The former asymmetry corresponds to unequal co- and contra-directional space-time weighted averaging~\cite{caloz2022GSTEM}, or Fresnel drag~\cite{huidobro2019fresnel}, while the latter asymmetry corresponds to different co- and contra-directional interference conditions due do the related Doppler red- and blue-shift.

\begin{figure}[h!]
\centering
\includegraphics[width=0.9\columnwidth]{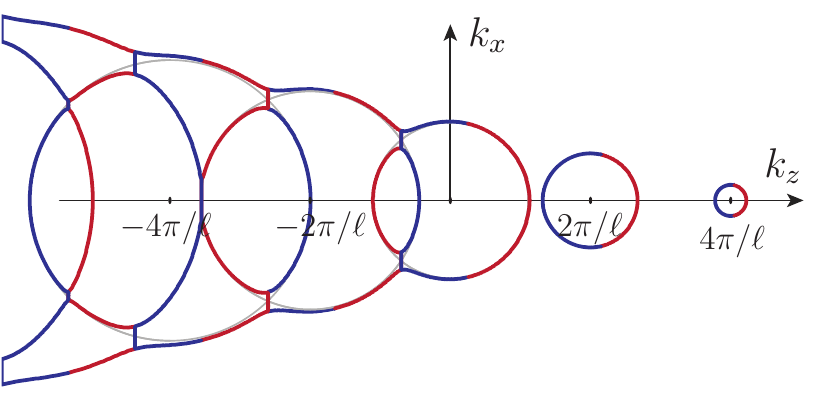}
\caption{Isofrequency diagram for the same crystal parameters as Fig.~\ref{fig:Dispersion} at $\omega^+=\omega^-=\omega_0=\pi$, with gray circles corresponding to empty-lattice approximation $(k_n+n 2\pi/\ell)=(\omega_0-n 2\pi/d)n_\tx{av}/c$~\cite{chutamir1972,mazor_oneway_2020}.}
\label{fig:iso}
\end{figure}

Figure~\ref{fig:iso} shows an isofrequency diagram, i.e., $\omega(k_z,k_x)$ contour curves, corresponding to an $\omega$-cut in the dispersion diagram of Fig.~\ref{fig:Dispersion}. This diagram has been obtained by the following procedure, varying this time the parameter $\theta_i^+$: 1)~find in \eqref{eq:disp_w} the values of $\omega_i^+$ that satisfy $\omega^+=\omega_0$ (red curves) and $\omega^-=\omega_0$ (blue curves); 2)~insert these values into~\eqref{eq:disp_kz} to obtain the corresponding $k_z^+$ (red) and $k_z^-$ (blue) values; 3)~use the formula $k_x=\omega_i^+n_i\sin\theta_i^+/c$ to determine the corresponding $k_x$ values; 4)~plot the corresponding $k_z^\pm(\omega_i^+,\theta_i^+)$ and $k_x(\omega_i^+,\theta_i^+)$ values.

\appendix

\section{Interface Matrix}\label{app:Interface_Matrix_Derivation}
The boundary conditions at an interface moving in the direction $z$ at the velocity $v$, read, in the case of s-polarization~\cite{bladel1984},
\begin{subequations}\label{eq:HyExcont}
\begin{equation}
   \left.E_{y1}-v B_{x1}=E_{y2}- v B_{x2}\right|_{z-v t=0},
\end{equation}
\begin{equation}
    \left.H_{x1}-v D_{y1}=
    H_{x2}-v D_{y2}\right|_{z-v t=0}.
\end{equation}
\end{subequations}
Inserting forward ($+$) and backward ($-$) plane waves with the fields $E_{yn}^\pm=E_n^\pm$, $D_{yn}^\pm=\epsilon_nE_{n}^\pm$, $H_{xn}^\pm=B_{xn}^\pm/\mu_n$ and \mbox{$cB_{xn}^\pm=\pm n_n E_{n}^\pm\cos\theta$} (with $n=1,2$) into these equations yields
\begin{subequations}\label{eq:HyDxcont}
\begin{equation}\label{eq:Hycont}
    \begin{split}
    E_{1}^+&(1-\beta n_1\cos\theta_1^+ )+E_{1}^-(1+\beta n_1\cos\theta_1^-)
 \\&=E_{2}^+(1-\beta n_2\cos\theta_2^+)+E_{2}^-(1+\beta n_2\cos\theta_2^-),
 \end{split}
\end{equation}
\begin{equation}
\begin{split}
E_{1}^+&(\cos\theta_1^+-\beta n_1)/\eta_1-E_{1}^-(\cos\theta_1^-+\beta n_1)/\eta_1\\&
=E_{2}^+(\cos\theta_2^+-\beta n_2)/\eta_2-E_{2}^-(\cos\theta_2^-+\beta n_2)/\eta_2,
\end{split}
\end{equation}
\end{subequations}
where we used $n_n=c\sqrt{\epsilon_n\mu_n}$ and $\eta_n=\sqrt{\mu_n/\epsilon_n}$. More compactly, these equations read
\begin{equation}
 \begin{bmatrix}
   f_1^+& f_1^-\\
  g_1^+ & -g_1^-
 \end{bmatrix}
 \begin{bmatrix}
    E_1^+ \\
    E_1^-
  \end{bmatrix}
  =
   \begin{bmatrix}
   f_2^+ & f_2^- \\
   g_2^+ & -g_2^-
 \end{bmatrix}
 \begin{bmatrix}
    E_2^+ \\
   E_2^-
  \end{bmatrix}.
\end{equation}
Isolating the column vector $E_2$ gives then Eqs.~(\ref{eq:Iji}). 

\section{Empty-Lattice Approximation}\label{app:empty-lattice}
The empty-lattice (or infinitesimal-perturbation) approximation is obtained by letting the refractive index and impedance contrast tend to zero, which corresponds to setting $A=1$ in \eqref{eq:A}. This transforms the spatial relation~(\ref{eq:disp_kz}) into
\begin{align}
\cos(k_z^\pm\ell\mp\Delta\phi^\tx{s}/2)
&=\cos\frac{\Sigma\phi^\tx{s}_i}{2}\cos\frac{\Sigma\phi^\tx{s}_j}{2}-\sin\frac{\Sigma\phi^\tx{s}_i}{2}\sin\frac{\Sigma\phi^\tx{s}_j}{2} \nonumber\\
& =\cos(\Sigma\phi^\tx{s}_i/2+\Sigma\phi^\tx{s}_j/2)
\end{align}
Setting the arguments of the first and last terms of this equation to be equal, isolating $k_z^\pm$ in the resulting relation, and substituting next \eqref{eq:def_barphis}, yields
\begin{equation}\label{eq:kz_av}
    k_z^\pm =\left(\Sigma\phi^\tx{s}_i+\Sigma\phi^\tx{s}_j\pm\Delta\phi^\tx{s}\right)/(2\ell)=\left(k_{zi}^\pm\ell_i+k_{zj}^\pm\ell_j\right)/(2\ell).
\end{equation}
Similarly, the temporal relation~\eqref{eq:disp_w} transforms to
\begin{equation}\label{eq:omega_av}
   \omega^\pm =\left(\omega_{i}^\pm d_i+\omega_{j}^\pm d_j\right)/(2d).
\end{equation}
Combining Eqs.~(\ref{eq:kz_av}) and~(\ref{eq:omega_av}) yields the empty-lattice approximation
\begin{equation}
   \frac{n_\tx{av}^\pm}{c}=
   \left(\frac{k_{zi}^\pm \ell_i+k_{zj}^\pm \ell_j}{\omega_i^\pm d_i+\omega_j^\pm d_j}\right)\frac{d}{\ell}.
\end{equation}







\bibliography{analysis_st_crystals}

\bibliographyfullrefs{analysis_st_crystals}


\end{document}